# Iterative optimization of photonic crystal nanocavity designs by using deep neural networks

TAKASHI ASANO[*] AND SUSUMU NODA

*Department of Electronic Science and Engineering, Kyoto University, Kyoto 615-8510 Japan*
*\*Corresponding author: tasano@kuee.kyoto-u.ac.jp*

**Abstract:**

Devices based on two-dimensional photonic-crystal (2D-PC) nanocavities, which are defined by their air hole patterns, usually require a high quality ($Q$) factor to achieve high performance. We demonstrate that hole patterns with very high $Q$ factors can be efficiently found by the iteration procedure consisting of: machine learning of the relation between the hole pattern and the corresponding $Q$ factor, and new dataset generation based on the regression function obtained by machine learning. First a dataset comprising randomly generated cavity structures and their first principles $Q$ factors is prepared. Then a deep neural network is trained using the initial dataset to obtain a regression function that approximately predicts the $Q$ factors from the structural parameters. Several candidates for higher $Q$ factors are chosen by searching the parameter space using the regression function. After adding these new structures and their first principles $Q$ factors to the training dataset, the above process is repeated. As an example, a standard silicon-based L3 cavity is optimized by this method. A cavity design with a high $Q$ factor exceeding 11 million is found within 101 iteration steps and a total of 8070 cavity structures. This theoretical $Q$ factor is more than twice of the previously reported record values of the cavity designs detected by the evolutionary algorithm and the leaky mode visualization method. It is found that structures with higher $Q$ factors can be detected within less iteration steps by exploring not only the parameter space near the present highest-$Q$ structure but also that distant from the present dataset.

## 1. INTRODUCTION

Photonic nanocavities based on artificial defects in two-dimensional (2D) photonic-crystal (PC) slabs **[1–11]** have received significant attention as structures that enable preservation of photons for extended times in small modal volumes. 2D-PC slab cavities are usually defined by defects in the triangular air hole lattice of the PC. For example, cavities can be defined by a defect consisting of three missing air holes (the so-called L3 cavity), a single missing hole (H0 cavity), or a line defect with a modulation of the lattice constants (heterostructure cavity). Photons of the cavity modes are confined in such nanocavities in the in-plane and vertical directions by Bragg reflection due to the air hole pattern of the 2D PC and total internal reflection due to the refractive index contrast between the PC slab and the surrounding air or cladding layers, respectively. We note that the in-plane reflection is usually almost perfect while the vertical reflection is only partial **[2]**. Thus, the total spectral intensity of the wavevector components that do not fulfill the total internal reflection condition, i.e., the leaky components, determines the cavity's quality ($Q$) factor **[12]**. So far, various methods of optimizing cavity designs with respect to the $Q$ factor have been proposed and demonstrated **[2–5, 12–19]**. Among them, the Gaussian envelope approaches **[2,3]**, the leaky position visualization approach **[17]**, and the analytic inverse problem approaches **[13,14]** utilize the knowledge of the physics of photon confinement mentioned above. For instance, the analytic inverse problem approaches are based on approximations that relate the cavities' structural parameters to the mode fields, and thus allow us to explicitly determine an optimized cavity geometry with less leaky components **[13,14]**. This type of approaches is very useful to optimize specific structural parameters, but targets are limited because suited analytical expressions are only available for certain cavity types. On the other hand, the Gaussian envelope and leaky position visualization approaches improve cavity designs based on the differences between the mode field calculated for the actual structure and the ideal mode field, which is artificially generated and has a minimum of leaky components **[2,3,17]**. The comparison of these fields enables identification of spatial positions where leakage of photons occurs. However, since these approaches cannot predict the optimized structure, the modifications required for a reduction of leakage have to be manually identified by trial and error. While these approaches are useful in early optimization stages, they cannot utilize the large degree of freedom that is inherent to the 2D geometry of the air hole pattern. The reports on optimization of 2D-PC nanocavity designs by these approaches have so far considered only up to nine structural parameters (e.g. symmetric displacements of certain holes) for optimization **[2,3,17]**, because it is difficult to manually locate better air hole patterns in the high-dimensional parameter space consisting of the positions of all individual air holes. Obviously, more systematic and automated methods of exploring high-dimensional parameter spaces are required to fully utilize the potential of 2D-PC nanocavities.

Minkov et al. utilized a genetic algorithm to explore the parameter space of the 2D-PC air hole pattern, and succeeded in tuning up to 11 parameters to find more suited nanocavity structures without using the physical knowledge of leaky components **[15,16,18]**. However, this approach requires a relatively large number of randomly generated sample cavity structures and their calculated $Q$ factors: they have reported that 100 cycles × 80 individuals = 8000 sample cavities (300 cycles × 120 individuals = 36000 sample cavities) were required to optimize five (seven) parameters in the L3 (H0) cavity **[16]**. The relatively large number of required sample cavities is considered to be a consequence of the genetic algorithm, which basically utilizes only the good cavities among the sample cavities generated in each cycle. Recently, we have proposed an approach based on deep learning, demonstrating optimization of 27 parameters of a heterostructure cavity using a training dataset consisting of 1000 randomly generated air hole patterns and their calculated $Q$ factors **[19]**. In **[19]**, we trained a neural network (NN) by the sample dataset to obtain an approximate function of the $Q$ factor with respect to the structural parameters. This regression function was then employed to detect new cavity structures that are likely to exhibit higher $Q$ factors. The important point is that not only high-$Q$ structures but also moderate or low-$Q$ structures can be useful when searching new cavity geometries with higher $Q$ factors (since both improve the accuracy of the regression function developed by the NN), although high-$Q$ sample cavity structures are of course more helpful. However, one problem of this approach is that structures with $Q$ factors much higher than that of the base cavity design are rarely generated during the preparation of the training dataset. Therefore, the accuracy of the regression function at the parameter space near extremely high $Q$ factors is low.

In this report, we propose an iterative optimization method to overcome this problem: here, the candidate structures for higher-$Q$ factors identified by the regression function at the present iteration step are added to the training dataset for the next step. The new dataset is used to derive an improved regression function. To increase the diversity of the new candidates, several different candidate-selection constraints are defined and their combinations are used to efficiently explore the parameter space. In order to avoid strong influences of initial discoveries, one constraint is that the new candidate should lie at a parameter space distant from the structures that have already been analyzed. Additionally, we employ several NNs that learn the dataset in different orders, resulting in different regression functions. With these we can partly account for the uncertainty of the prediction by a NN. By repeating the optimization cycles, cavity structures that are important for detection of high-$Q$ cavity structures are automatically accumulated in the dataset. To demonstrate this, we optimize the design of a silicon (Si) L3 cavity via 25 parameters. We are able to detect a structure with a maximum $Q$ factor of almost 11 million by generating a total of 8070 sample structures within 101 iterations. This theoretical $Q$ factor is more than two times larger than the $Q$ factors of Si-based L3 cavity structures found by the genetic algorithm **[16]** and leaky mode visualization approaches **[17]**.

## 2. FRAMEWORK

In this section we explain the procedures of the proposed iterative optimization method, which contains the preparation, learning, structure search, validation, and dataset update phases. The latter four phases are repeated to iteratively improve the regression function developed in the learning phase and the following structure search. The general design of the preparation and learning phases can be found in **[19]**. First of all, we assume that the type of 2D-PC cavity that is to be optimized, is known (in Section 3 we choose the L3 cavity). Next, the preparation phase consisting of the following three procedures (I) to (III) has to be implemented:

  I. Select the structural parameters of the base cavity (such as air hole positions and radii) that should be considered for optimization. Generate many sample cavity structures by randomly varying the selected parameters within a certain meaningful range.

  II. Calculate the $Q$ factors of the sample cavities generated in (I) by a first principles method to obtain the training dataset consisting of the sample cavity structures and the corresponding $Q$ factors.

  III. Prepare deep NNs that have input nodes corresponding to the structural parameters selected in (I) and have a single output node corresponding to the $Q$ factor.

The learning phase is described by the following procedure:

  IV. Train the deep NNs prepared in (III) to learn the relation between the structure and the $Q$ factor using the dataset prepared in (I) and (II) (only for the first round) or the updated dataset obtained in (VII) (for the following rounds). Let each deep NN learn the dataset in a different order so that they acquire different approximation functions of $Q$.

The structure search phase consists of

  V. Starting from a randomly chosen initial cavity structure, gradually change the structural parameters using the gradient (in the parameter space) of the approximated $Q$ factor that is predicted by a trained deep NN. By this process, one new candidate structure with a potentially higher $Q$ factor is located. Various candidate structures are prepared by using different deep NNs and by applying different constraints (described later).

The validation phase is straightforward:

  VI. Determine the accurate $Q$ factors of the candidate structures by a first principles calculation.

After the learning, structure search and validation phases, the training dataset is updated and the next iteration cycle is carried out as follows:

VII. Add the sets of the structures obtained in (V) and the *Q* factors calculated in (VI) to the training dataset.

VIII. Go to (IV).

By repeating the procedures (IV)–(VII), the sample cavities that are important for locating high-*Q* structures are automatically accumulated, because both correct and wrong predictions constitute important information for the development of an improved regression function. Figure 1 briefly illustrates the concept of the approach for optimization explained above.

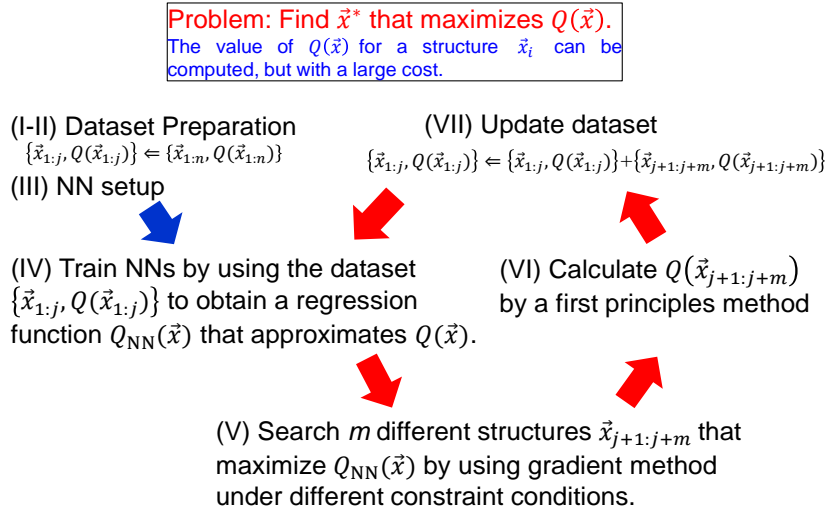

Fig. 1. The iterative optimization proposed in this paper. Each cavity with the structural parameters selected in the preparation phase is represented by a unique high-dimensional vector $\vec{x}$. The accurate *Q* as a function of $\vec{x}$, $Q(\vec{x})$, can be calculated by first principle approaches but is costly to compute. $\vec{x}_{i:j}$ denotes the set of structures and consists of $\vec{x}_i, \vec{x}_{i+1}, \ldots, \vec{x}_j$. $Q(\vec{x}_{i:j})$ denotes the corresponding set of *Q* factors, and $\{\vec{x}_{1:j}, Q(\vec{x}_{1:j})\}$ is used to refer to the dataset consisting of the chosen cavity structures and their *Q* factors. The number of sample structures in the initial dataset is *n*. $Q_{NN}(\vec{x})$ represents a low-cost regression function that approximates $Q(\vec{x})$ and is obtained by training a neural network (NN) using the training dataset $\{\vec{x}_{1:j}, Q(\vec{x}_{1:j})\}$. $Q_{NN}(\vec{x})$ is only used to locate new structures via the gradient, but the values are not explicitly discussed in this work.

## 3. Optimization of the cavity design for a Si-based L3 nanocavity

In this section, we demonstrate the optimization of the cavity design for a L3 cavity made of Si by the proposed iterative optimization. The results are useful for device development and also provide a benchmark for the optimization performance of the presently used algorithm. The numbers given can be compared with those in previously reported methods **[16,17]**, because the Si-based L3 nanocavity is a standard 2D-PC nanocavity.

### 3.1. Preparation phase

Figure 2 shows the basic structure of the presently considered L3 nanocavity, where the lattice constant is *a*, the radius of each air hole is 0.25 *a*, the thickness of the slab is 0.5366 *a*, and the refractive index of the slab material (Si) is *n* = 3.46. These values were chosen by considering the standard dimensions of fabricated nanocavities (*a* = 410 nm, *t* = 220 nm) operating at optical communication wavelengths **[9,20]**, and the refractive index of Si at these wavelengths. The radii of the air holes are the same as those used in **[16]**, and the slab thickness is similar to that in **[16]** (0.55 *a*). The color plot in Fig. 1 shows the electric field distribution of the fundamental mode in the *y*-direction ($E_y$). The distribution was calculated for the base cavity structure by a first principle method [three-dimensional finite-difference time-domain (3D-FDTD) method], and the resulting *Q* factor of the base structure is 7160. The modal volume $V_{cav}$ of the mode is 0.61 cubic wavelength in the material $(\lambda/n)^3$. Further details of the calculation conditions are provided in **[19]**.

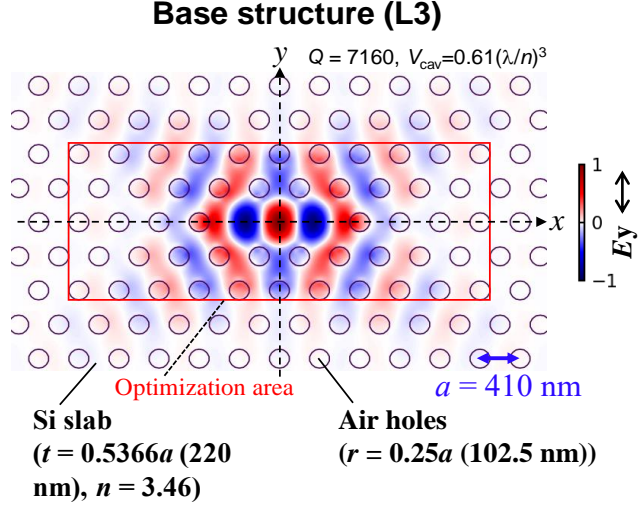

Fig. 2. A three-missing-air-holes (L3) cavity is used as the base structure for structural optimization. The lattice constant $a$ is 410 nm. The circles indicate the air holes (hole radius: 102.5 nm = 0.25 $a$) formed in Si slab with a refractive index $n$ = 3.46 and a thickness of 220 nm (0.5366 $a$). The distribution of the y-component of the electric field ($E_y$) of the fundamental resonant mode is plotted in color. The theoretical $Q$ factor and modal volume $V_{cav}$ of the base structure determined by FDTD are 7160, and 0.61 $(\lambda/n)^3$, respectively. The displacements of the 50 air holes inside the red square are the structural parameters that are used to optimize the cavity design with respect to $Q$.

[Step (I)]: The positions of the 50 air holes within the area of 11 ($a$) × 5 (rows) (indicated by the red square in Fig. 2) are the structural parameters that are used to optimize the cavity design with respect to the $Q$ factor, because most of the electric field intensity of the mode concentrates in this area [19]. Each sample cavity structure (labelled by index $i$) is defined by the base structure and a set of 2D displacement vectors $\{\vec{d}_1, \vec{d}_2, ...\}_i$, where $\vec{d}_h = (d_{hx}, d_{hy})$ defines the displacement of $h$-th air hole in the x–y plane and $h$ enumerates all air holes that are selected for structural optimization (from 1 to 50 in the present case). The parameter space vector of structure $i$, $\vec{x}_i$ as defined in Fig. 1, is a single column vector with the structure $(d_{1x}, d_{1y}, d_{2x}, …, d_{50y})^\mathsf{T}$ and contains displacements corresponding to the single set $\{\vec{d}_1, \vec{d}_2, ..., \vec{d}_{50}\}_i$. Although we have 100 degrees of freedom in the 2D displacements of 50 air holes, the actual degrees of freedom in the present analysis are 25 because we have to impose mirror symmetries with respect to the central x and y axes to obtain high $Q$ factors [12].

[Step (II)]: Random displacements are applied to all air holes in the x- and y-directions in such a way that the mirror symmetries of the structure are maintained and that a uniform distribution between -0.1$a$ to 0.1$a$ is obtained. The appropriate magnitude of the fluctuation has been determined in previous manual optimizations of L3 cavities [2, 17]. In this demonstration, we initially prepare $n$ = 1000 random nanocavity structures (the whole set is denoted by $\vec{x}_{1:n}$) using the above outlined displacement restrictions, and calculate their $Q$ factors using the 3D-FDTD method. The obtained set of $Q$ values, $Q(\vec{x}_{1:n})$, exhibits a distribution between $10^3$ and $10^5$, and the average is 6700. Because the first principles $Q$ values of the initial set are spread over two orders of magnitudes, and this difference should increase in the subsequent optimization cycles, we employ $\log_{10}(Q(\vec{x}_i))$ as the target of machine learning. As a result, the initial training dataset consists of the structural parameters $\vec{x}_i$ and $\log_{10} Q(\vec{x}_i)$ of the 1000 structures, i.e., $\{\vec{x}_{1:n}, \log_{10} Q(\vec{x}_{1:n})\}$ instead of $\{\vec{x}_{1:n}, Q(\vec{x}_{1:n})\}$.

[Step (III)]: 10 four-layer-NNs with the same configuration as in [19] are prepared. The input nodes are two-channel 2D tables, where each channel corresponds to the x- and y-components of $\{\vec{d}_1, \vec{d}_2, ..., \vec{d}_{50}\}$. The first layer is a convolutional layer [21] with 50 filters with a size of 3 (holes) × 5 (rows) × 2 (channels) that is connected to the second layer with 450 units. The last part of each NN comprises the third layer (200 units), the fourth layer (50 units), and the output layer (one unit). These layers are fully connected through rectified linear units (ReLU [22]) and affine transformations. Stochastic information selection units (DROPOUT [23]) are additionally inserted between the third and fourth layer. The single output unit is intended to predict $\log_{10}(Q(\vec{x}))$.

### 3.2. Learning phase

[Step (IV)]: For this phase, we employ a conventional loss function $L$ consisting of two terms: the squared difference between the output of the NN and the teacher data (i.e., $\log_{10}(Q(\vec{x}_i))$), and the summation of the squared connection weights $w_m$ in the network (weight decay method [24]), where the latter is used to avoid the overfitting,

$$L = |Output(i) - \log_{10} Q(\vec{x}_i)|^2 + \frac{1}{2}\lambda \sum_m w_m^2. \qquad (1)$$

For the hyperparameter λ we use 0.00333 determined from the (10-fold) cross-validation method. In the training process, we randomly select one set $\{\vec{x}_i, \log_{10} Q(\vec{x}_i)\}$ from the training dataset $\{\vec{x}_{1:j}, \log_{10} Q(\vec{x}_{1:j})\}$, where $j$ is the number of samples in the present dataset as defined in Fig. 1, and change the internal parameters of the NN to reduce $L$ using the back-propagation method [25]. Here, the actual output of the NN is referred to as $\log_{10} Q_{\text{NN}}$, where $Q_{\text{NN}}$ is an approximation of the $Q$ factor. We apply the momentum optimization method to speed up convergence [26], where the learning rate and the momentum decay rate are set to $1.0\times10^{-4}$ and 0.9, respectively. The random selection of one structure and following reduction of $L$ by using the back-propagation method is repeated $5\times10^4$ times. 10 separate NNs are trained by the same method, but with different orders of data feeding. Therefore, after the training, each NN has acquired different internal parameters, which widens the divergence of the candidate structures that are generated in the following step (V).

### 3.2. Structure search phase

[Step (V)]: Several candidate structures (here, we use $m = 70$) with potentially higher $Q$ factors are generated using the gradient method. For this we define the loss function $L'$,

$$L' = \left|\log_{10} Q_{\text{target}} - \log_{10} Q_{\text{NN}}\right|^2 + (Artifitial\ Loss), \qquad (2)$$

and calculate the gradient of $L'$ with respect to $\vec{x}$ (i.e., $\nabla_{\vec{x}} L'$) using the back-propagation method [25], where $Q_{\text{target}}$ is set to a very high value (here, we use $1.0\times10^8$). Starting from a randomly generated initial structure defined in the parameter space by $\vec{x}_k^{ini}$ ($k > j$), we incrementally change the structure to reduce the loss $L'$ (i.e., $\vec{x}_k \leftarrow \vec{x}_k + \Delta \vec{x}$, where $\Delta \vec{x}$ is a set of incremental hole displacements calculated from $\nabla_{\vec{x}} L'|_{\vec{x}_k}$ based on the momentum method [26]). The artificial loss term in Eq. (2) is used to constrain the structural parameter space that is explored during the optimization, and different conditions are used to obtain different candidate structures. We designed the following three types of artificial losses, where $\lambda'$ is a control parameter.

(A) Squared distance from the base structure or the best structure in the previous round:

$$\frac{1}{2}\lambda'|\vec{x} - \vec{x}_b|^2, \qquad (3)$$

Where $\vec{x}_b$ refers to the sample structure with the highest $Q$ in the previous rounds (i.e., the highest $Q$ among $Q(\vec{x}_{1:j})$). (In the case of the first round, $\vec{x}^b$ is set to zero, because the base structure has no displacements). This artificial loss is designed to explore the parameter space in the vicinity of the best structure in the previous rounds.

(B) Squared distance from a randomly generated initial structure $\vec{x}_k^{ini}$:

$$\frac{1}{2}\lambda'\left|\vec{x} - \vec{x}_k^{ini}\right|^2. \qquad (4)$$

This artificial loss forces exploration of unknown parameter space stochastically. It is expected that a structure with a higher $Q$ that is not predictable from the training dataset, can be accidentally found by using this artificial loss.

(C) Sum of the inverse of the distances from all the structures in the training data set:

$$\lambda' \sum_{i \leq j} |\vec{x} - \vec{x}_i|^{-1}. \tag{5}$$

This artificial loss increases as the parameter space vector of the structure that is being optimized approaches the locations of the known structures, $\vec{x}_i$ with $i \leq j$. This restriction forces exploration of unknown parameter space more strictly than (B).

For the present demonstration, we designed and investigated the following three strategies of candidate generation:

**Strategy (A):** Each NN generates 7 different candidates using the artificial loss (A) with 7 different $\lambda'$ (3.0, 1.0, 0.1, 0.01, 0.001, 0.0001, 0.00001).

**Strategy (A+B):** Each NN generates 3 candidates using the artificial loss (A) with 3 different $\lambda'$ (1.0, 0.1, 0.01), 1 candidate without using artificial losses ($\lambda' = 0$), and 3 different candidates using the artificial loss (B) with 3 different $\lambda'$ (1.0, 0.1, 0.01).

**Strategy (A+C):** Each NN generates 3 candidates using the artificial loss (A) with 3 different $\lambda'$ (1.0, 0.1, 0.01), 1 candidate without using artificial losses ($\lambda' = 0$), and 3 different candidates using the artificial loss (C) with 3 different $\lambda'$ (1.0, 0.1, 0.01).

*3.3. Validation and update phases*

[Step (VI)]: The $Q$ factors of the 70 candidate structures obtained in step (V) for each strategy are determined by 3D-FDTD calculations. The calculation conditions are the same as in **[19]**.

[Step (VII)]: The new data consisting of 70 candidate structures (defined by $\vec{x}_{j+1:j+70}$) and their $Q$ factors ($Q(\vec{x}_{j+1:j+70})$) calculated in (VI) for each strategy are added to each strategy's training dataset.

[Step (VIII)]: Steps (IV) to (VII) are repeated 101 times. During this iterative optimization of the regression function $Q_{\mathrm{NN}}(\vec{x})$ and, consequently, also that of the cavity design, different series of training datasets are accumulated for each strategy, and 8070 sample cavities are accumulated in each dataset after 101 rounds of optimization.

*3.4. Results*

Figure 3 shows the highest $Q$ factors of the additional 70 samples structures generated in each iteration step cycle as a function of the size of the used training dataset. The corresponding $Q_{\mathrm{NN}}$ are not discussed in the following, because the regression function is only employed to identify structures with potentially higher $Q$ factors (via the gradient method). The results for the different strategies (A), (A+B), and (A+C) are shown with the blue, orange, and green curves, respectively. We find that the highest $Q$ achieved in each round overall increases with further iteration although some fluctuations exist. The highest $Q$ factors of the structures that have been detected by 100 iterations of cavity design optimization are $5.75 \times 10^6$, $9.12 \times 10^6$, and $1.10 \times 10^7$ for strategies (A), (A+B), (A+C), respectively. Figure 4 plots the inter-structure distances between the best structure in the present round and the best structure in the previous rounds in terms of the parameter space vector $\vec{x}$, indicating how large the modifications in each round of optimization are. It can be confirmed that the inter-structure distances tend to decrease as the optimization proceeds. The inter-structure distances for strategy (A+C) is basically always larger than those for the other structures, and that for strategy (A+B) is larger than that for (A) only at early stages (< 4000 samples). The air hole displacements of the structures with the highest $Q$ factors found during 100 optimization cycles for the three strategies are shown in Fig. 5. The distribution of $E_y$ and the modal volume $V_{\mathrm{cav}}$ of the cavity mode are shown as well. It is interesting to note that the displacements of the best cavities for the three strategies are significantly different. $V_{\mathrm{cav}}$ of the optimized cavities are 0.73, 0.68, and 0.74 $(\lambda/n)^3$ for strategies (A), (A+B), (A+C), respectively, which are slightly larger than that of the base structure (0.61 $(\lambda/n)^3$, Fig. 2).

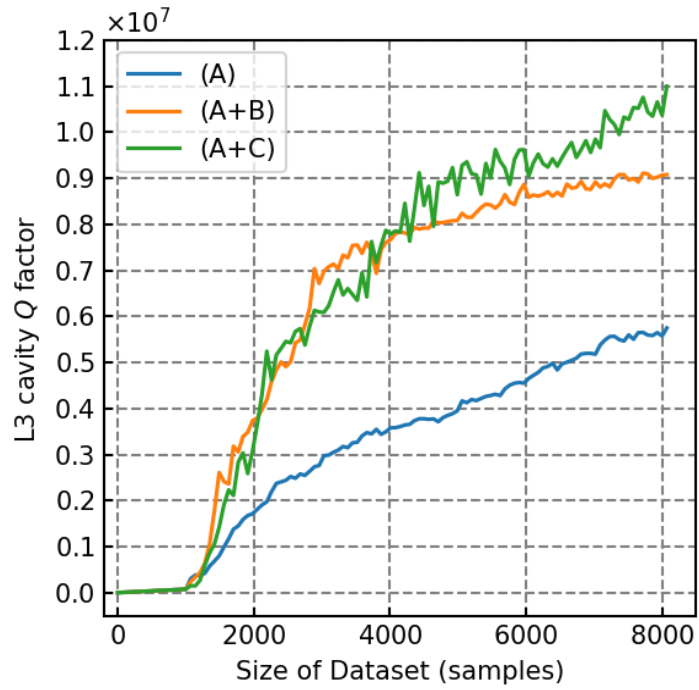

Fig. 3. The highest $Q$ factor of the additional 70 sample cavities generated in one round as a function of in the size of the used training dataset. The results for the three different strategies (A), (A+B), and (A+C) are shown with blue, orange, and green curves, respectively. The highest $Q$ factors of the candidates identified in 101 rounds of optimization of the regression function are $5.75 \times 10^6$, $9.12 \times 10^6$, and $1.10 \times 10^7$ for strategies (A), (A+B), and (A+C), respectively.

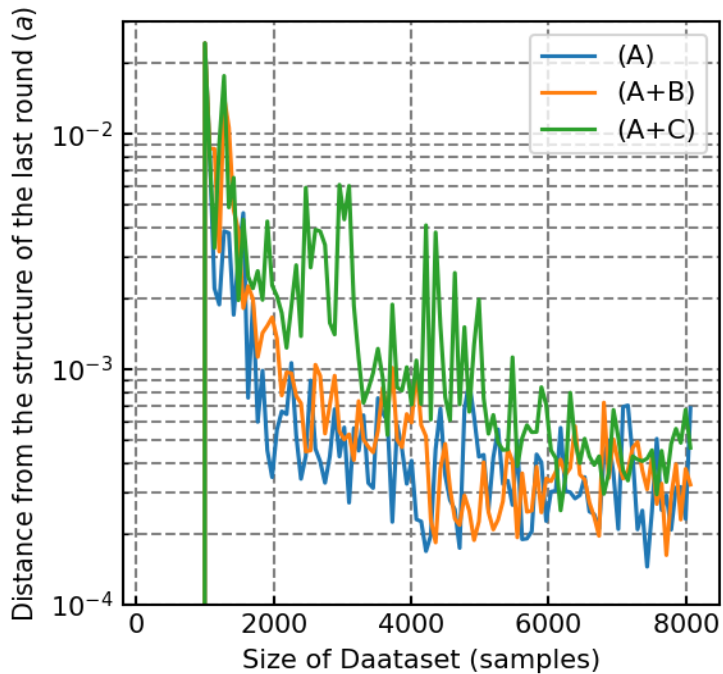

Fig. 4. Inter-structure distances between the structure with the highest $Q$ factor in the present round (at $\vec{x}_b'$ in the parameter space) and that in the previous rounds (at $\vec{x}_b$) as a function of the size of the used training dataset. The inter-structure distance $|\vec{x}_b' - \vec{x}_b|$ indicates how large the modifications in each round of optimization are. The results for the three different strategies (A), (A+B), and (A+C) are shown with blue, orange, and green curves, respectively.

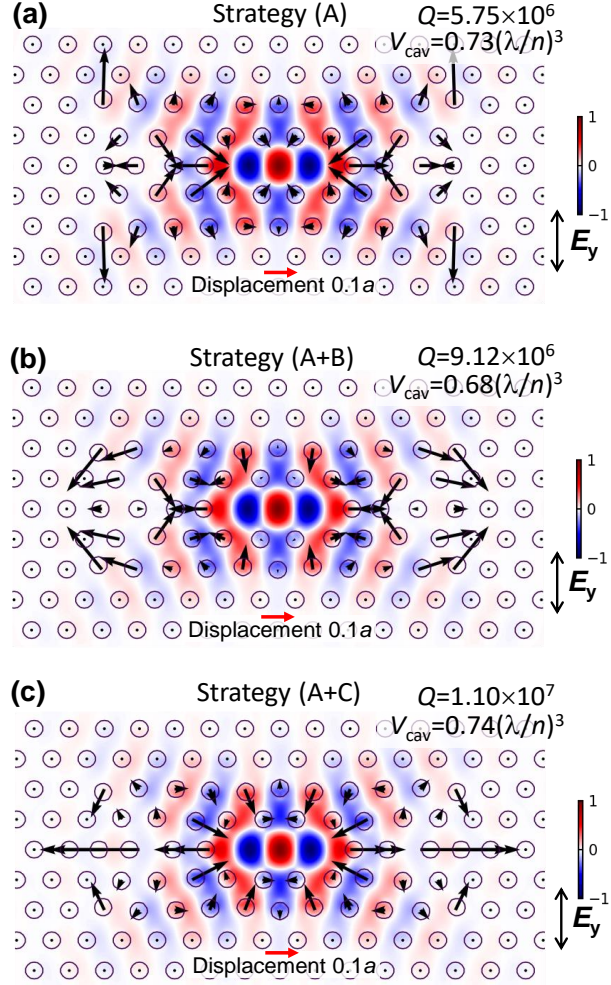

Fig. 5. Air hole displacements of the optimized cavities with the highest Q factors found during the entire optimization for (a) strategy (A), (b) strategy (A+B), and (c) strategy (A+C). The circles represent the determined optimum positions of the air holes, and the arrows represent the displacement vectors with the scale shown by the red arrow. The electric field distributions of $E_y$ are also plotted in color. The highest $Q$ factors of the structures generated by our proposed method during 101 iteration cycles are $5.75 \times 10^6$, $9.12 \times 10^6$, and $1.10 \times 10^7$ for strategies (A), (A+B), (A+C), respectively.

## 4. DISCUSSION

### 4.1. Performance of the three strategies

At first, we compare the results of the iterative optimization proposed in this report with the optimization results of the previously reported NN-based optimization method [19], which corresponds to the results obtained in the first round. After the first round of optimization, 1070 samples cavities had been accumulated, and the highest $Q$ factors were $3.09 \times 10^5$, $2.44 \times 10^5$, and $1.57 \times 10^5$ for strategies (A), (A+B), (A+C), respectively. The improvements relative to the $Q$ of the base structure (7160, Fig. 2) are about 43, 34, 22 times for (A), (A+B), (A+C), respectively. The iterative optimization method was able to detect structures with much higher $Q$ factors: the highest $Q$ factors that have been found until the final 101th round are $5.75 \times 10^6$, $9.12 \times 10^6$, and $1.10 \times 10^7$, and thus the improvement ratios are about 800, 1270, 1540 times for (A), (A+B), (A+C), respectively. This evidences that the proposed method is very effective compared to the previous method, because computation costs for first principles calculations increased only by 8 times (from 1070 to 8070 sample cavities).

In the following we compare the three strategies with respect to the best $Q$ and the computation costs. The best $Q$ factors found with strategies (A+B) and (A+C) are about 1.6 and 1.9 times larger than that of (A). The differences in the improvement ratios originate from the differences in the methods of generating candidate cavity structures. In strategy (A), candidates for high $Q$ structures are generated by exploring the parameter space following the gradient of $Q_{NN}$ predicted by the trained NN while keeping the distance from the best structure in the previous rounds small. This constraint is controlled by parameter $\lambda'$, which was changed from 3 to $1\times10^{-5}$ to generate 7 different candidates. Candidates with higher $Q$ factors are frequently generated, but the candidate structures are generated according to the past experience (although some randomness is introduced by the initial structures). Therefore, the possibility to get stuck in a local maximum during the repetition of the optimizations is relatively large. The rapid decrease of the inter-structure distance for this strategy shown in Fig. 4 supports this interpretation.

In strategy (A+B), half of the candidates are generated according to the past experience, and half of the candidates are generated by exploring the parameter space near randomly generated initial structures. The latter approach is expected to add diversity to the generated candidates and the training dataset. It is important to note that the latter half is not just a random generation; here, candidates are explored based on experience (the gradient of $Q_{NN}$) while the space to be explored is intentionally limited. This can prevent getting stuck in an already known local maxima. This explanation is supported by the larger inter-structure distances for this strategy compared to those for strategy (A) in the early stages of optimization (Fig. 4; < 4000 samples). It seems that the advantage of this strategy decreases as the number of iteration cycles increases as shown in Figs. 3 and 4. We explain this with the higher probability of an overlap between the randomly generated initial structure and some sample in the training dataset after many iterations, reducing the diversity of the generated candidates. However, this strategy detected a structure with a $Q$ factor that is 1.6 times larger than the best structure found with strategy (A). It is noted that the computation cost for this strategy is the same as that for (A), because the computation cost for the artificial loss (B) [Eq. (4)] equals that of the artificial loss (A) [Eq. (3)].

In strategy (A+C), half of the candidates are generated according to the past experience, and half of the candidates are generated by exploring the parameter space according to the gradient of $Q_{NN}$ while avoiding the space near the already known structures in the training data set. Therefore, unknown parameter space is explored more explicitly compared to the case of using artificial loss (B). The maximum $Q$ factor found with this strategy is 20% larger than that detected with strategy (A+B) as shown in Figs. 3 and 5. Moreover, the tendency to detect significantly higher $Q$ factors in the next iteration step is still not saturated even at 100 optimization cycles (Fig. 3; the inclination of the green curve is relatively steep). This is in contrast to the case of (A+B), and means that strategy (A+C) can avoid local maxima more effectively. The much larger inter-structure distances of this strategy shown in Fig. 4 support this interpretation. The drawback is the increase in the computation cost: as can be seen from Eq. (5), the evaluation cost for the artificial loss (C) scales with the number of samples in the training dataset ($N$) while the other terms in the loss function do not scale with $N$ as can be seen in Eqs. (2)–(4). However, the evaluation cost of the artificial loss (C) scales much slower than that of the Bayesian optimization discussed later.

*4.2. Comparison with other optimization methods*

Here, we compare the L3 cavity optimization performances of our proposed method and other state-of-the-art optimization methods as a benchmark. The genetic algorithm based method was used in [16] to optimize 5 parameters in the Si-based L3 cavity and enabled detection of a structure with a $Q$ of 4.2 million by using ~8000 sample cavities. Reference [17] optimized 9 parameters in L3 cavity using a leaky component visualization method, and found a structure with a $Q$ of 5.3 million by using 200 sample cavities. In comparison, our proposed method was used to optimize 25 parameters of the L3 cavity and we detected a structure with a $Q$ of 11.0 million by using 8070 sample cavities generated with strategy (A+C). The maximum $Q$ detected by the proposed method is more than 2.6 times larger than that found in [16], while the number of samples cavities that have been used are almost the same.

Compared to the leaky component visualization method, the $Q$ obtained in the present method is more than two times larger. Although the number of sample cavities used in [17] is only about 200, these sample structures had to be explored manually, which usually consumes similar time and more effort compared to the proposed automated method. Therefore, our proposed method has provided a structure that is a more optimized than those of the two previous methods while the computation costs are similar. We consider that the higher optimization efficiency of the proposed method has two origins: a training database that contains all experiences accumulated during the whole calculation, and also the aggressive search of unknown parameter space, which results in generation of candidate structures that are useful for the optimization.

*4.3. Comparison with Bayesian optimization*

Finally, we compare the proposed method and the well-known Bayesian optimization in the context of generic optimization methods. The Bayesian optimization is a powerful tool to optimize a black-box function that is expensive to calculate [27,28]. In this method, an approximate function (usually a so-called Gaussian process [28]) that predicts not only the mean but also variance (uncertainty) of the values of the black-box function is generated by using the present dataset (the training dataset in our case). Then, an acquisition function that evaluates the probability of obtaining better values is prepared based on the predicted mean and uncertainty. As a new observation point (= candidate), the point with the highest value of the acquisition function is searched in its parameter space, and the value of the black-box function at this observation point is calculated and added to the dataset. This procedure is iterated many times. We note that the approach of our proposed method constitutes almost the same procedure. As explained in Section 2, our framework uses a NN to construct an approximation function of the black-box function $Q(\vec{x})$. The artificial loss (C) roughly evaluates the inverse of the uncertainty, and the use of several NNs trained with different data feeding orders also corresponds to the evaluation of the uncertainty of the predicted values.

However, there are important differences between the Bayesian optimization and our proposed optimization method. One is the computation cost for the search of candidate structures in the high-dimensional parameter space (this applies to the so-called normal Bayesian optimization only), and the other is the computation cost of the learning of a large-scale training dataset (this also applies to other types of Bayesian optimization). Concerning the former difference, the normal Bayesian optimization usually uses direct search methods without relying on derivatives [28], and therefore sufficient search in a high-dimensional parameter space is impossible from the viewpoint of computation cost (a practical parameter space is usually limited to less than 10 dimensions) [28–30]. The gradient method starting from random initial points would be useful for the search in high-dimensional space, but it is difficult to implement because the gradient of the acquisition function in the normal Bayesian optimization tends to be zero over a wider region in the parameter space as its dimension increases, because of the characteristics of the kernel function [29]. The Bayesian optimization with the elastic Gaussian process [29] can overcome this issue, but computation costs for the learning of large-scale datasets remain high as discussed later. The random embedding method [30] can also treat high-dimensional parameter spaces but only under the rather restrictive assumption that the numbers of important dimensions are very small. In contrast, our method is able to effectively utilize the gradient method in high-dimensional space because the gradient of the loss function does not disappear owing to the RELU nonlinear layers in the NNs and the properly designed artificial loss terms. Therefore, a parallel trial of gradient-based searches starting from many randomly generated initial points works well in our case.

Regarding the relationship between the scale of the training dataset ($N$) and the computation cost for the training, the cost in the Bayesian optimization scales with $N^3$ because the inverse of a $N \times N$ matrix has to be calculated for the training [28]. (Ref. [31] utilized a deep neural network with a Bayesian linear regressor in the last hidden layer to resolve this issue. However, the maximum feasible dimension of the parameter space is still limited because direct (parallel) search is utilized [31].) Fortunately, the training cost of a NN only scales with $N^1$, and thus a large-scale dataset can be employed to increase the precision of the candidate search, which is especially important for the optimization in a high-dimensional parameter space. In addition, our method employs the gradient method starting from random initial points, which enables efficient exploration of a high-dimensional space. In total, it is considered that the proposed approach benefits from the characteristics of a NN-based regression, which enables training of a large-scale dataset and search in a high-dimensional parameter space while introducing the policy of the Bayesian optimization (i.e., the mean and variation of the prediction are taken into account).

## 5. CONCLUSION

We proposed and demonstrated a new approach for optimizing 2D-PC nanocavity designs, which have large degrees of structural freedom. This approach comprises the repetition of the following four steps: training of NNs to learn the relationship between cavity structure and the $Q$ factor using the present dataset, generation of candidate structures using the trained NNs, calculation of their $Q$ factors, and finally adding the new structures and $Q$ factors to the dataset. The key point of this approach is to generate a variety of candidate structures to avoid getting stuck in a local maximum in the high-dimensional parameter space. For this purpose, we prepared several NNs and trained them with different data feeding orders. In addition, we designed three artificial loss terms and used them to generate candidate structures by employing the regression function provided by a trained NN. It was demonstrated that the artificial loss term that increases near the known structures in the dataset works most efficiently to increase the speed of generating structures with higher $Q$-factors: this method generated an optimized Si-based L3 nanocavity structure with a $Q$ factor of 11 million (here, 25 parameters were fine-tuned using 101 iterations and a total of 8070 sample cavities). This $Q$ factor is more than 2 times larger than the $Q$ factors obtained by previously reported methods, while computation costs and efforts are similar. We also compared our method and the Bayesian optimization in the context of generic optimization methods. The proposed approach is effective not only for the optimization of 2D-PC nanocavity designs, but also for generic optimization problems in high-dimensional parameter space.

**Funding**

JSPS KAKENHI (19H02629); A project commissioned by the New Energy and Industrial Technology Development Organization (NEDO).

**Acknowledgments**

The authors would like to thank Mr. Koki Saito for his helpful textbook on deep learning written in Japanese (Deep learning from scratch, O'Reilly Japan).

**REFERENCES**

1. S. Noda, A. Chutinan, and M. Imada, "Trapping and emission of photons by a single defect in a photonic bandgap structure," Nature **407**, 608–610 (2000).
2. Y. Akahane, T. Asano, B.-S. Song, and S. Noda, "High-Q photonic nanocavity in a two-dimensional photonic crystal," Nature **425**, 944-947 (2003).
3. B. S. Song, S. Noda, T. Asano, and Y. Akahane, "Ultra-high-Q photonic double-heterostructure nanocavity," Nat. Mater. **4**, 207-210 (2005).
4. T. Asano, B.-S. Song, and S. Noda, "Analysis of the experimental Q factors (~ 1 million) of photonic crystal nanocavities," Opt. Express **14**(5), 1996–2002 (2006).
5. E. Kuramochi, M. Notomi, S. Mitsugi, A. Shinya, T. Tanabe, and T. Watanabe, "Ultrahigh- Q photonic crystal nanocavities realized by the local width modulation of a line defect," Appl. Phys. Lett. **88**, 041112 (2006).
6. Y. Takahashi, H. Hagino, Y. Tanaka, B. S. Song, T. Asano, and S. Noda, "High-Q nanocavity with a 2-ns photon lifetime," Opt. Express **15**, 17206-17213 (2007).
7. E. Kuramochi, H. Taniyama, T. Tanabe, A. Shinya, and M. Notomi, "Ultrahigh-Q two-dimensional photonic crystal slab nanocavities in very thin barriers," Appl. Phys. Lett. **93**, 111112 (2008).
8. Z. Han, X. Checoury, D. Néel, S. David, M. El Kurdi, and P. Boucaud, "Optimized design for $2\times10^6$ ultra-high Q silicon photonic crystal cavities," Opt. Commun. **283**, 4387-4391 (2010).
9. H. Sekoguchi, Y. Takahashi, T. Asano, and S. Noda, "Photonic crystal nanocavity with a Q-factor of ~9 million," Opt. Express **22**, 916-924 (2014).
10. T. Asano, Y. Ochi, Y. Takahashi, K. Kishimoto, and S. Noda, "Photonic crystal nanocavity with a Q factor exceeding eleven million," Opt. Express **25**, 1769-1777 (2017).
11. T. Asano, S. Noda, "Photonic Crystal Devices in Silicon Photonics," *Proc. IEEE* **106**, 1–13 (2018).
12. K. Srinivasan and O. Painter, "Momentum space design of high-Q photonic crystal optical cavities," Opt. Express **10**, 670–684 (2002).
13. D. Englund, I. Fushman, and J. Vucković, "General recipe for designing photonic crystal cavities," Opt. Express **13**, 5961–5975 (2005).
14. Y. Tanaka, T. Asano, and S. Noda, "Design of Photonic Crystal Nanocavity With Q-Factor of ~ $10^9$," J. Light. Technol. **26**, 1532–1539 (2008).
15. Y. Lai, S. Pirotta, G. Urbinati, D. Gerace, M. Minkov, V. Savona, A. Badolato, and M. Galli, "Genetically designed L3 photonic crystal nanocavities with measured quality factor exceeding one million," Appl. Phys. Lett. **104**, 241101 (2014).
16. M. Minkov and V. Savona, "Automated optimization of photonic crystal slab cavities," Sci. Rep. **4**, 5124 (2015).
17. T. Nakamura, Y. Takahashi, Y. Tanaka, T. Asano, and S. Noda, "Improvement in the quality factors for photonic crystal nanocavities via visualization of the leaky components," Opt. Express **24**, 9541-9549 (2016).

18. M. Minkov, V. Savona, and D. Gerace, "Photonic crystal slab cavity simultaneously optimized for ultra-high Q / V and vertical radiation coupling," Appl. Phys. Lett. **111**, 131104 (2017).
19. T. Asano, S. Noda, "Optimization of photonic crystal nanocavities based on deep learning," Opt. Express. **26**, 32704–32716 (2018).
20. K. Maeno, Y. Takahashi, T. Nakamura, T. Asano, S. Noda, "Analysis of high-Q photonic crystal L3 nanocavities designed by visualization of the leaky components," *Opt. Express* **25**, 367-376 (2017).
21. Y. LeCun, B. Boser, J. S. Denker, D. Henderson, R. E. Howard, W. Hubbard, and L. D. Jacklel, "Handwritten Digit Recognition with a Back-Propagation Networks," in Proceedings of Advances in Neural Information Processing Systems, pp. 396-404 (1990).
22. X. Glorot, A. Bordes, and Y. Bengio: "Deep Sparse Rectifier Neural Networks," in Proceedings of Artificial Intelligence and Statistics, pp. 315-323 (2011).
23. N. Srivastava, G. Hinton, A. Krizhevsky, I. Sutskever, and R. Salakhutdinov, "Dropout: A Simple Way to Prevent Neural Networks from Overfitting," Journal of Machine Learning Research **15**, 1929–1958 (2014).
24. A. Krogh and J. A. Hertz, "A Simple Weight Decay Can Improve Generalization," in Proceedings of Advances in Neural Information Processing Systems, pp. 950–957 (1991).
25. D. E. Rumelhart, G. E. Hinton, and R. J. Williams, "Learning representations by back-propagating errors," Nature **323**, 533–536 (1986).
26. B. T. Polyak, "Some methods of speeding up the convergence of iteration methods," USSR Computational Mathematics and Mathematical Physics **4**, 791-803 (1964).
27. D. R. Jones, M. Schonlau, and W. J. Welch. "Efficient global optimization of expensive black-box functions," J. of Global Optimization **13**, 455– 492 (1998).
28. B. Shahriari, K. Swersky, Z. Wang, R. P. Adams, N. De Freitas, "Taking the human out of the loop: A review of Bayesian optimization," Proc. IEEE **104**, 148–175 (2016).
29. S. Rana, C. Li, S. Gupta, V. Nguyen, S. Venkatesh, "High Dimensional Bayesian Optimization with Elastic Gaussian Process," *Proc.* of the 34th *Int. Conf. on Mach. Learn.* **70**, 2883-2891 (2017).
30. Z. Wang, F. Hutter, M. Zoghi, D. Matheson, N. De Freitas, "Bayesian optimization in a billion dimensions via random embeddings," *J. Artif. Intell. Res.* **55**, 361–367 (2016).
31. J. Snoek et al., "Scalable Bayesian Optimization Using Deep Neural Networks," Proc. of the 32nd International Conference on Machine Learning **37**, (2015). (available at http://arxiv.org/abs/1502.05700).